\documentstyle[prl,twocolumn,aps,epsf]{revtex}
\newcommand{\bo}{\={o} }

\begin{document}

\title{Thermodynamically Important Contacts in Folding\\ of Model
Proteins}

\author{Antonio Scala$^1$\thanks{Present address: INFM, Udr Rome ``La
Sapienze,'' Piazzele Aldo Moto 2, I-00185, Rome, Italy.}, Nikolay
V. Dokholyan$^{1,2}$\footnote{Corresponding author. Email:
dokh@wild.harvard.edu}, Sergey V. Buldyrev$^1$ and H. Eugene
Stanley$^1$}

\address{$^1$Center for Polymer Studies and Department of Physics,
	  Boston University, Boston, MA 02215\\ $^2$Department of
	  Chemistry and Chemical Biology, Harvard University,
	  Cambridge, MA 02138 }

\date{\today}
\maketitle

\vspace{5pt}

\begin{abstract}

We introduce a quantity, the entropic susceptibility, that measures the
thermodynamic importance---for the folding transition---of the contacts
between amino acids in model proteins. Using this quantity, we find that
only one equilibrium run of a computer simulation of a model protein is
sufficient to select a subset of contacts that give rise to the peak in
the specific heat observed at the folding transition. To illustrate the
method, we identify thermodynamically important contacts in a model
46-mer. We show that only about 50\% of all contacts present in the
protein native state are responsible for the sharp peak in the specific
heat at the folding transition temperature, while the remaining 50\% of
contacts do not affect the specific heat.

\end{abstract}

\pacs{PACS number(s): 87.15.-v, 87.15.Aa, 87.15.Cc}

\vspace{30pt}

Proteins are heteropolymers, composed of 20 types of amino acids, that
perform specific functions. The amino acid composition of proteins
determines their unique structure, function, and folding
kinetics. Understanding the relevance of the interactions between amino
acids to protein folding is a complex task that has been the subject of
a number of theoretical and experimental studies
\cite{Anifsen73,Go83,Bryngelson89,Dill90,Karplus94,Shakhnovich96,%
Klimov96,Shakhnovich97,Li97,Dokholyan98,Maritan98,Dokholyan00a,%
Dokholyan00xxx}.  The transition from the unfolded to the folded state
of a protein is accompanied by a drastic reduction of the entropy. In
one popular scenario, the folding transition for short proteins is
analogous to the nucleation process at a first order transition
\cite{Go83,Shakhnovich96,Shakhnovich97,Dokholyan00a}, with competition
between two free energy minima: the folded state with low energy and
entropy and the unfolded state with high energy and entropy. These
two minima are separated by a free energy barrier corresponding to the
transition states.

At the folding transition temperature, $T_f$, there is an abrupt change
in the energy of the system resulting in a pronounced peak in the
specific heat. At $T_f$, a small increase in interaction energy
$\epsilon_{ij}$ between amino acids $i$ and $j$ (``contact strength'')
results in rapid transition to the folding state, while a small decrease
in contact strength results in transition to the unfolded
state. However, different amino acids have a different contribution to
the folding transition.  Small variation in $\epsilon_{ij}$ for
different pairs $i$ and $j$ has a different effect on the folding
transition. Here, we study the thermodynamic importance of each
interaction during folding by computing the entropic
susceptibility---the response function to a small perturbation of
$\epsilon_{ij}$.

We assume that the protein potential energy is additive in the pair
potentials (contacts)
\begin{equation}
U \equiv \frac{1}{2} \sum_{i,j} U_{ij} \equiv \frac{1}{2}\sum_{i,j}
 \epsilon_{ij} \phi(\vec{r}_i,\vec{r}_j)\, ,
\label{eq:H}
\end{equation} 
where $U_{ij}$ is the energy of a single pair, $\phi(\vec{r}_i,
\vec{r}_j)$ models the shape of the potential and protein at positions
$\vec{r}_i$ and $\vec{r}_j$.  We define the entropic susceptibility,
$\chi_{ij}$, of a contact between amino acids $i$ and $j$ as
\begin{equation}
\chi_{ij} \equiv 
\left. -\epsilon_{ij} \frac{\partial S}{\partial \epsilon_{ij} } \right.
\equiv
\beta^2 \left( 
	\langle U ~ U_{ij} \rangle - 
	\langle U \rangle \langle U_{ij}\rangle 
	\right)
\equiv \beta^2 \langle \delta U ~ \delta U_{ij} \rangle \, ,
\label{eq:chi}
\end{equation}
where $\delta U \equiv U - \langle U \rangle$, $\delta U_{ij} = U_{ij} -
\langle U_{ij} \rangle$, and $\langle\dots \rangle$ is the Boltzmann
average~\cite{Footnote1}.

\begin{figure}[htb] 
\narrowtext \centerline{
\hbox  {
        \vspace*{0.5cm}
        \epsfxsize=10.0cm
        \epsfbox{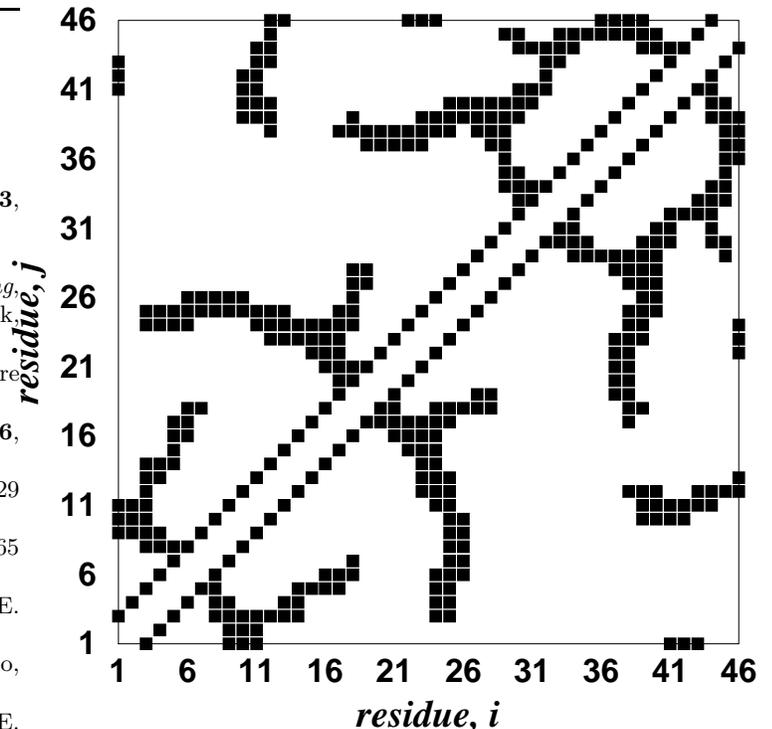}
       \hspace*{0.3cm}
       }
          }     
\caption{Contact map of the native state of the 46-mer: dark squares
denote residues that have contacts in the native state. Interactions
are assigned according to G\bo model \protect\cite{Go83}: all pairs of
residues that have a contact in the native state are assigned
attractive potential ($\epsilon_{ij}=-1$), while remaining pairs of
residues are assigned to a repulsive potential ($\epsilon_{ij}=+1$).}
\label{cmap} 
\end{figure}

The entropic susceptibility measures the effect of a contact strength
perturbation on the folding transition of the protein, thus identifying
the thermodynamic relevance of such contact for the folding
transition. Next, we demonstrate how this measure can be used to study
contributions of the various contacts between amino acids in the protein
for the folding transition. We simulate the ``beads on a string''
protein model~\cite{Dokholyan00a}, where the amino acids are hard
spheres of unit mass, with the centers at the positions of the
corresponding $\alpha$-carbons. The potentials of interaction between
amino acids are square wells of depth $\epsilon_{ij}$. We study the
46-mer (the folding transition temperature is at $T_f\approx 1.44$) that
has been examined in \cite{Dokholyan00a}. We use G\bo model for the
contact potential, $U_{ij}$: $U_{ij}$ is attractive ($\epsilon_{ij}=-1$)
if the contact exists in the native (ground) state, otherwise the
contact potential is repulsive ($\epsilon_{ij}=+1$)
\cite{Footnote2,Go81}. Our simulations employ the discrete MD algorithm
and are performed using methods described in
\cite{Dokholyan98,Dokholyan00a,DMD}. The matrix of native contacts of
the 46-mer is shown in Fig.~\ref{cmap}. This particular 46-mer is known
to have a stable native state and to undergo first-order-like folding
$\rightleftharpoons$ unfolding transitions without stable intermediates
\cite{Dokholyan00a}.

\begin{figure}[htb] 
\narrowtext \centerline{
\hbox  {
        \vspace*{0.5cm}
        \epsfxsize=10.0cm
        \epsfbox{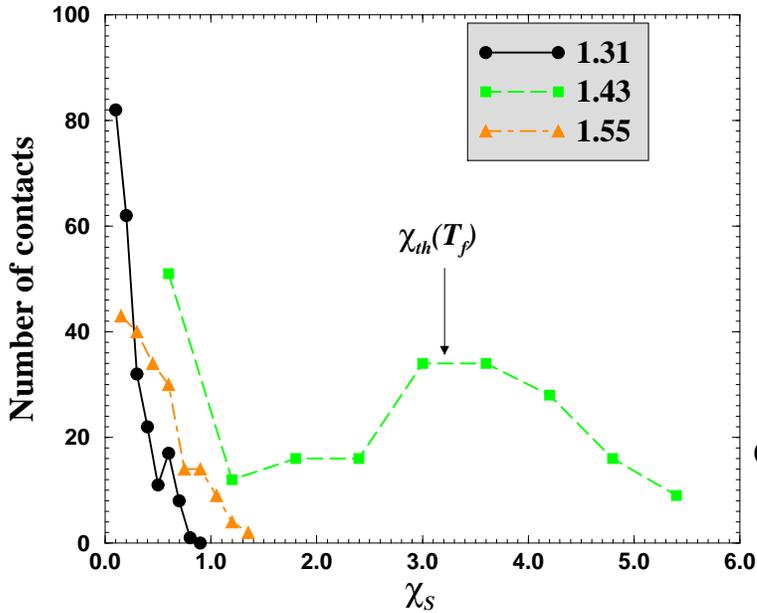}
       \hspace*{0.3cm}
       }
          }     
\caption{Histogram of the entropic susceptibility, $\chi_{ij}$, for the
46-mer at temperatures $T=1.31$, 1.44, 1.55.  At $T_f$ the distribution
of $\chi_{ij}$ has a pronounced peak, centered at
$\chi_{ij}=3.2$. Accordingly, we choose $\chi_{\mbox{\scriptsize
th}}=3.2$.}
\label{chishyst} 
\end{figure}

We calculate $\chi_{ij}$ at different temperatures below and above
$T_f$. A histogram of the values of $\chi_{ij}$ for various $T$ is shown
in Fig.~\ref{chishyst}. For $T \approx T_f$ the distribution has a
pronounced peak at large values of $\chi_{ij}$, which indicates that
there is a separation of all contacts in two distinct sets with large
and small values of $\chi_{ij}$. The set of contacts with large values
of $\chi_{ij}$ are ``thermodynamically important contacts,'' since for
these contacts a small variation in their strength is correlated with a
drastic change in the entropy of the model protein. To select the
thermodynamically important contacts, we define a temperature-dependent
threshold $\chi_{\mbox{\scriptsize th}}(T)$ corresponding to the value
of $\chi_{ij}$ where the distribution has a maximum in the space of all
contacts.

\begin{figure}[htb] 
\narrowtext \centerline{
\hbox  {
        \vspace*{0.5cm} \epsfxsize=10.0cm \epsfbox{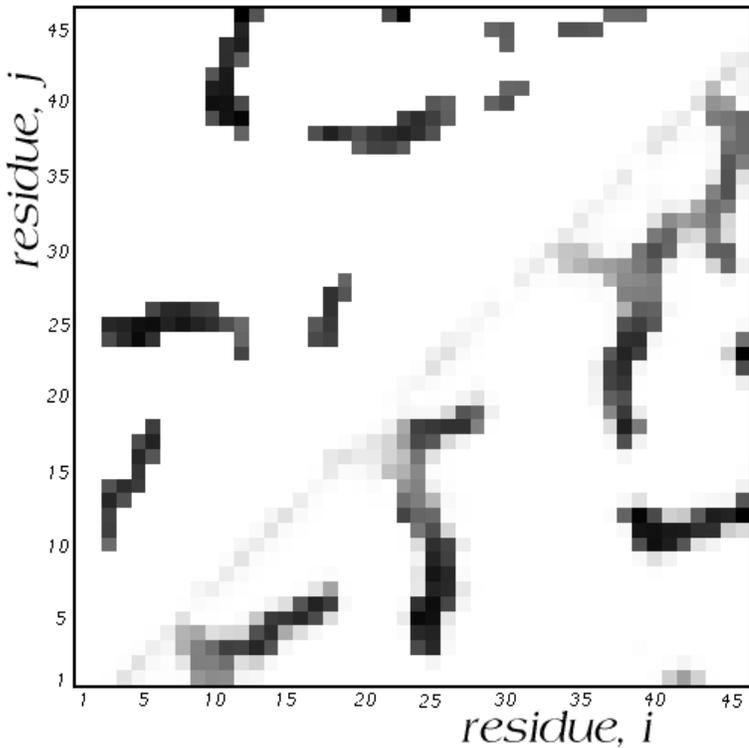}
        \hspace*{0.3cm} } }
\caption{The role of the thermodynamically important contacts: the lower
corner is the intensity map of the entropic susceptibility, $\chi_{ij}$,
obtained from the simulations of the 46-mer at $T_f$. Darker colors
correspond to higher values of $\chi_{ij}$.  The upper corner is the
``filtered'' map, where only values of $\chi_{ij}$ above the threshold
$\chi_{\mbox{\scriptsize th}}=3.2$ defined in
Fig.~\protect\ref{chishyst} are presented. Note that short ranged
contacts ($i\approx j$, corresponding to near-to-the-diagonal elements
of the matrix $\chi_{ij}$) do not contribute significantly to the change
of entropy at the folding transition, while the relevant long-ranged
contacts ($|i-j|\gg 1$) are clustered in the islands in the filtered
map. Specifically, nucleic contacts determined in
\protect\cite{Dokholyan00a} belong to the cluster in the top left corner
with $i\approx 10$ and $j\approx 40$.}
\label{chismap}
\end{figure}

Interestingly, thermodynamically important contacts are not randomly
distributed in 3d space but are rather concentrated within well-defined
structural regions in a model protein. Figure~\ref{chismap} represents
the intensity map of the values $\chi_{ij}$. In the upper part of
Fig.~\ref{chismap}, we show only the values of $\chi_{ij}$ that are
above the threshold $\chi_{\mbox{\scriptsize th}}(T_f)=3.2$ which,
according to our definition, corresponds to the thermodynamically
important contacts.  Although 50\% of the contacts are above threshold,
the filtered map of Fig.~\ref{chismap} shows that they are clustered
together and are among well-defined regions of the model
protein. Further, we find that the regions of thermodynamically
important interactions ($\chi_{ij}(T) > \chi_{\mbox{\scriptsize
th}}(T)$) in the filtered map remain qualitatively the same as the ones
shown in Fig.~\ref{chismap} for temperatures in the range $T=T_f\pm5\%$.

\begin{figure}[htb] 
\narrowtext \centerline{
\hbox  {
        \vspace*{0.5cm}
        \epsfxsize=10.0cm
        \epsfbox{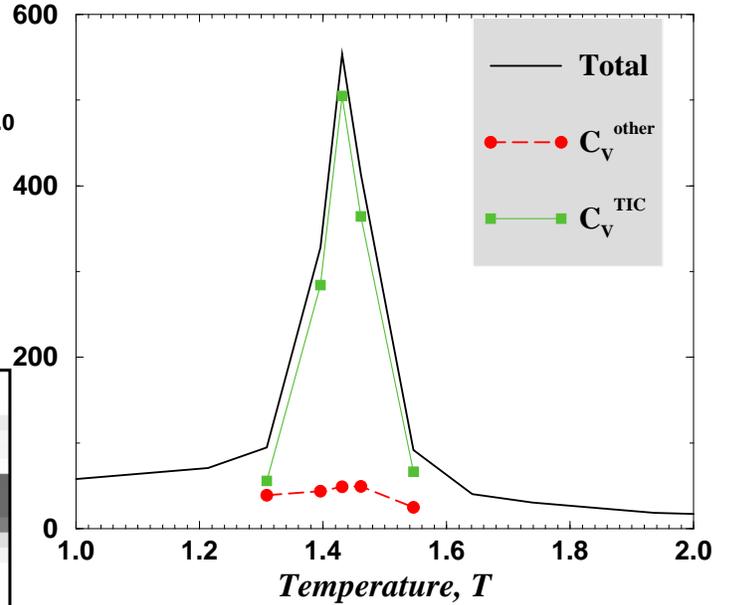}
       \hspace*{0.3cm}
       }
          }     
\caption{Specific heat as a function of temperature of the 46-mer
(solid line). There is a pronounced peak in the distribution of
$\chi_{ij}$ in the region $1.31<T<1.55$. We separate the contribution to
the specific heat from the thermodynamically important contacts
$C_V^{\mbox{\scriptsize TIC}}$ (squares) and the remaining ones
$C_V^{\mbox{\scriptsize other}}$ (circles).
The number of contacts above threshold is always found to be
approximatively $50\%$ of the number of native contacts.}
\label{cvcompare} 
\end{figure}

To verify that the thermodynamically important contacts are indeed
thermodynamically the most relevant to the folding of our $46$-mer, we
measure the contribution of thermodynamically important contacts to the
specific heat
\begin{equation}
C_V \equiv  \frac{1}{2} \sum_{ij} \chi_{ij} \, .
\label{eq:cv}
\end{equation}
Thus, we can interpret $\chi_{ij}$ as the contribution to $C_V$ of a
single contact. It is then possible to partition $C_V$ as
\begin{equation}
C_V= C_V^{\mbox{\scriptsize TIC}}+C_V^{\mbox{\scriptsize others}}\, ,
\label{eq:2cv}
\end{equation}
where $C_V^{\mbox{\scriptsize TIC}}$ arises from the thermodynamically
important contacts, and $C_V^{\mbox{\scriptsize others}}$ from contacts
below the threshold $\chi_{\mbox{\scriptsize
th}}(T)$. Fig.\ref{cvcompare} shows that the thermodynamically important
contacts give a sharp contribution to the specific heat around $T_f$. We
find the number of contacts above threshold $\chi_{\mbox{\scriptsize
th}}(T_f)$ is about $50\%$ of the number of contacts in the native
state, in agreement with Flory-type arguments~\cite{Dokholyan98}.

It is natural to inquire whether the thermodynamically important
contacts could be determined by analyzing the average contact energies
$\langle U_{ij} \rangle$, which are related to the contact frequency map
\cite{Maritan98}. For square well potentials, $\langle U_{ij} \rangle =
\epsilon_{ij} \langle f_{ij} \rangle$ where $f_{ij}$ is the contact
frequency for amino acids $i$ and $j$. We find that the contacts with
the largest values of $f_{ij}$ are nearest and next nearest
neighbors. Thus, in order to account for the long-range contacts we have
to go beyond the estimation of frequencies. An alternative way of
computing the entropic susceptibility is to note that the contact
frequencies are related to the change in free energy $F$
\begin{equation}
 \left. -\epsilon_{ij} \frac{\partial F}{\partial \epsilon_{ij} }
 \right.  \equiv \beta \langle U_{ij} \rangle
\end{equation}
and therefore the entropic susceptibility can be rewritten as
\begin{equation}
\chi_{ij} \equiv -\epsilon_{ij} \frac {\partial }{\partial
\epsilon_{ij} } \left(-\frac{\partial F}{\partial T }\right) = -
\frac{\partial \beta \langle U_{ij}\rangle}{\partial T}
={1\over T^2}\langle U_{ij}\rangle-{\partial\langle
U_{ij}\rangle\over\partial T}{1\over T}.
\end{equation}
Thus, the information about the thermodynamically important contacts can
be inferred from the temperature {\it derivative\/} of the frequency
map.

\begin{figure}[htb]
\narrowtext \centerline{
\hbox  {
        \vspace*{0.5cm}
        \epsfxsize=10.0cm
        \epsfbox{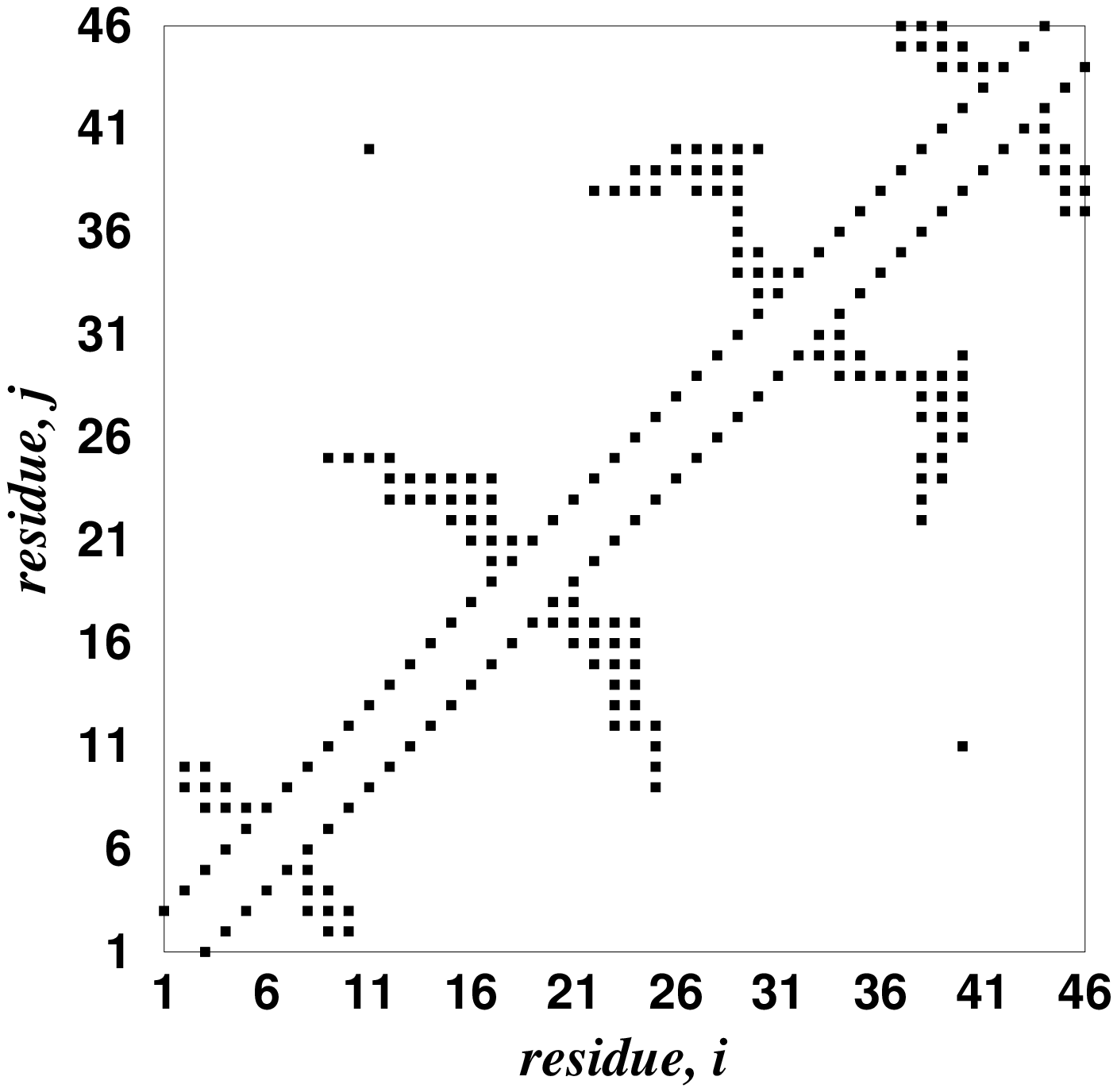}
       \hspace*{0.3cm}
       }
          }
\caption{The map of core contacts of the 46-mer: the dark squares denote
contacts between any two residues $i$ and $j$ with frequencies
$f_{ij}>0.5$. The core is comprised mostly of local contacts.}
\label{fig:core}
\end{figure}

``Core contacts'' were defined in \cite{Dokholyan98} as those that
form most stable elements of the protein three-dimensional structure
that remains intact at folding transition temperature. Specifically,
they were defined as contacts that are present with frequency above
0.5 at $T_f$. Molecular dynamics simulations performed at $T=T_f$ (see
Fig.~\ref{fig:core}) show that these contacts are mostly
short-range. (The range for the contact between residues $i$ and $j$
is defined as $|i-j|$). This result is not surprising since local
contacts can form with high probability even in the unfolded state at
$T\approx T_f$.

In contrast, we find that the thermodynamically important contacts are
mostly long-range, for which $|i-j|\gg 1$ (see Fig.~\ref{chismap}).
According to our definition, the thermodynamically important contacts
correlate with the potential energy, thus, they are likely to be present
in the folded state with the low potential energy and are likely to be
absent in the unfolded states with high potential energy. Therefore, we
believe that they are important for stabilization of the native
structure. This hypothesis is in agreement with the general observation
\cite{Abkevich95,Govindarajan95} that long-range interactions are
important for protein stabilization.

We also find that the set of the thermodynamically important contacts
contain all five nucleic contacts ($(11,39)$, $(10,40)$, $(11,40)$,
$(10,41)$, and $(11,41)$) discovered in \cite{Dokholyan00a},
indicating the dual role some amino acids play in protein folding: the
nucleic residues, which play crucial role in the kinetics of folding
transition, may also be important for stabilizing proteins in their
native state. The evidence for the existence of such residues is
supported by evolutionary \cite{Dokholyan00xxx,Dokholyan00xxVar} and
phenomenological studies \cite{Plaxco98,Fersht00,Dokholyan00yy}.

In conclusion, we demonstrate that by calculating the cross correlations
between the potential energy of a single contact and the total potential
energy of a model protein, it is possible to identify the set of contacts
that are thermodynamically most relevant to the folding process. The
tool of identifying thermodynamically important contacts is simple and
can be implemented in the molecular dynamics studies of model
proteins. The computational effort can be directed to aid experimental
studies of real proteins.

\subsection*{acknowledgements}

We thank F. Sciortino, E. I. Shakhnovich, and M. Vendruscolo for very
useful discussions.  NVD is supported by NIH postdoctoral fellowship
1F32 GM20251-01. The Center for Polymer Studies acknowledges the
support of the NSF.



\begin{thebibliography}{99}

\bibitem{Anifsen73} 
C.~B. Anifsen, Science {\bf 181}, 223 (1973).

\bibitem{Go83} 
N. G\bo, Ann. Rev. Biophys. Bioeng. {\bf 12}, 183 (1983).

\bibitem{Bryngelson89} 
J. D. Bryngelson and P. G. Wolynes, J. Phys. Chem. {\bf 93}, 6902
(1989).

\bibitem{Dill90} 
K.~A. Dill, Biochemistry {\bf 29}, 7133 (1990).

\bibitem{Karplus94} 
M. Karplus and E.~I. Shakhnovich, in {\it Protein Folding}, edited by
T. Creighton (W. H. Freeman, New York, 1994), pp.~127--196.

\bibitem{Shakhnovich96} 
E. I. Shakhnovich, V.~I. Abkevich and O. Ptitsyn, Nature {\bf 379}, 96
(1996).

\bibitem{Klimov96} 
D.~K. Klimov and D. Thirumalai, Phys. Rev. Lett.  {\bf 76}, 4070 (1996).

\bibitem{Shakhnovich97} 
E. I. Shakhnovich, Curr. Opinion Struc. Biol. {\bf 7}, 29 (1997).

\bibitem{Li97} 
H. Li, C. Tang, N.~S. Wingreen, Phys. Rev. Lett.  {\bf 79}, 765 (1997).

\bibitem{Dokholyan98} 
N. V. Dokholyan, S. V. Buldyrev, H. E. Stanley, and E. I. Shakhnovich,
Folding \& Design {\bf 3}, 577 (1998).

\bibitem{Maritan98} 
C. Micheletti, J. R. Banavar, A. Maritan, and F. Seno, Phys. Rev. Lett.
{\bf 82}, 3372 (1998).

\bibitem{Dokholyan00a} 
N. V. Dokholyan, S. V. Buldyrev, H. E. Stanley, and E. I. Shakhnovich,
J. Mol. Biol. {\bf 296}, 1183 (2000).

\bibitem{Dokholyan00xxx} 
N. V. Dokholyan, L. A. Mirny, and E. I. Shakhnovich, preprint
(cond-mat/0007084).

\bibitem{Footnote1} 
In general, the potential energy can take the form $U \equiv \sum_n
\sum_{i_1,...,i_n} U^{(n)}_{i_1,...,i_n}$, where $U^{(n)}_{i_1,...,i_n}$
is the $n$-body interaction between amino acids $i_1,...,i_n$. The
resulting expression for the entropic susceptibility is defined then as
$\chi^{(n)}_{i_1,...,i_n}= \beta^2 \langle \delta U ~ \delta
U^{(n)}_{i_1,...,i_n} \rangle$.

\bibitem{Footnote2} 
The repulsive potential is necessary to ensure the existence of a non
degenerate ground (folded) state.

\bibitem{Go81} 
N. G\bo and H. Abe, Biopolymers {\bf 20}, 991 (1981); H. Abe and
N. G\bo, Biopolymers {\bf 20}, 1013 (1981).

\bibitem{DMD} 
B. J. Alder and T. E. Wainwright, J. Chem. Phys. {\bf 31}, 459 (1959);
A. Yu. Grosberg and A. R. Khokhlov, {\it Giant Molecules\/} (Academic
Press, Boston, 1997) [see Appendix]; M. P. Allen and D. J. Tildesley,
{\it Computer Simulation of Liquids\/} (Clarendon Press, Oxford, 1987),
Ch.~3; D. C. Rapaport, {\it The Art of Molecular Dynamics Simulation\/}
(Cambridge University Press, Cambridge, 1997), Ch.~12.

\bibitem{Abkevich95} 
V.~I. Abkevich, A.~M. Gutin, and E.~I. Shakhnovich, J. Mol. Biol. {\bf
252}, 460 (1995).

\bibitem{Govindarajan95} 
S. Govindarajan and R. A. Goldstein, Biopolymers {\bf 36}, 43 (1995).

\bibitem{Dokholyan00xxVar} 
N.~V. Dokholyan and E. I. Shakhnovich, in {\it Proceedings of the
International School of Physics ``Enrico Fermi'' Course CXLV: Protein
Folding, Evolution and Design\/} (Varenna, Italy, 2000), submitted.

\bibitem{Plaxco98} 
K. W. Plaxco, K.~T. Simons, and D. Baker, J. Mol. Biol.  {\bf 277}, 985
(1998).

\bibitem{Fersht00} 
A.~R. Fersht, Proc. Natl. Acad. Sci. US {\bf 97},
1525 (2000).

\bibitem{Dokholyan00yy} 
N.~V. Dokholyan, L. Li and E. I. Shakhnovich, {\it
Proc. Natl. Acad. Sci. US}, submitted.

\end{thebibliography}
\end{document}